\documentclass[%
aps,
pre,
twocolumn,
superscriptaddress,
amsmath,
amssymb,
citeautoscript
]{revtex4-1}
\usepackage{graphicx}% Include figure files
\usepackage{dcolumn}% Align table columns on decimal point
\usepackage{bm}% bold math
\usepackage{hyperref}% add hypertext capabilities
\usepackage{comment}
\usepackage{xcolor}
\usepackage{soul}
\usepackage[normalem]{ulem}
\usepackage{cleveref}

\newcommand{\sidiff}{S1}
\newcommand{\sifit}{S2}

\begin{document}

\preprint{}

\title{
Junctional-Fluctuation-Mediated Fluidisation of\\ Multi-Phase Field Epithelial Monolayers
}

\author{James N. Graham}
\affiliation{Rudolf Peierls Centre for Theoretical Physics, University of Oxford, Oxford OX1 3PU, United Kingdom}

\author{Jan Rozman}
\email{jan.rozman@physics.ox.ac.uk}
\affiliation{Rudolf Peierls Centre for Theoretical Physics, University of Oxford, Oxford OX1 3PU, United Kingdom}
\date{\today}

\begin{abstract}
We analyse a multi-phase field model for an epithelial monolayer with pairwise adhesions between neighbouring cells following an Ornstein-Uhlenbeck process, representing the stochastic turnover of junctional molecular motors. These fluctuations in junctional adhesion result in rearrangements in the tissue, fluidising it and producing diffusive cell motion. Similar junctional fluctuations have proven a very useful tool in the vertex model literature, and we hope they will be equally helpful to the multi-phase field model approach. Moreover, we observe that the cells' effective diffusion coefficient depends non-monotonically on the persistence time of the fluctuations, confirming results previously observed in the vertex model. 
\end{abstract}

\maketitle

%||||||||||||||||||||||||||||||||||||||||||||||||||||||||||||
%||||||||||||||||||||||||||||||||||||||||||||||||||||||||||||
%||||||||||||||||||||||INTRODUCTION||||||||||||||||||||||||||
%||||||||||||||||||||||||||||||||||||||||||||||||||||||||||||
%||||||||||||||||||||||||||||||||||||||||||||||||||||||||||||
\section{Introduction}
The physics of epithelial cells and tissues, their rearrangements, and their motion has been of long interest~\cite{alert:2020,xi2019material,shankar2022topological}. Epithelia have been studied using a range of theoretical and computational approaches such continuum models~\cite{ranft2010fluidization,serra2012mechanical,saw2017topological}, particle-based models~\cite{szabo2006phase,sepulveda2013collective,garcia2015physics}, and cellular Potts models~\cite{graner1992simulation,glazier1993simulation,hirashima2017cellular}.

One extensively used approach to modelling epithelial monolayers are 2D vertex models~\cite{honda1980much,farhadifar2007influence,fletcher2014vertex,bi2015density,alt2017vertex}. In this formalism, cells are represented as polygons tiling the surface, with their vertices moving in response to forces. One avenue of research in vertex models has been understanding the role of junctional tension fluctuations, as would arise from the stochastic turnover of junctional molecular motors, in the rearrangement and motion of cells. This was modelled by the junctional tensions in the vertex model following an Ornstein-Uhlenbeck process~\cite{curran:2017,krajnc:2020,duclut:2021,yamamoto:2022}: such tension fluctuations can induce T1 transitions, thereby fluidising the tissue. Notably, Ref.~\cite{yamamoto:2022} reported that the effective diffusion coefficient of cells in the resulting fluidised tissue depends non-monotonically on the persistence time of the tensions.

Fluctuating tensions in the vertex model formalism have proven a very useful tool and have been used in studies of, e.g., wound healing~\cite{tetley2019tissue}, cell sorting~\cite{krajnc:2020}, budding in three dimensions~\cite{rozman:2020,rozman:2021}, nematic forces with a global direction~\cite{duclut:2021a}, and long-range order in tissues~\cite{keta2025long}. Therefore, in this paper, we adapt junctional fluctuations to another cell-resolution approach to modelling tissues, multi-phase field models~\cite{alert:2020, nonomura:2012, najem:2016, mueller:2019, mouregomez:2019, camley:2017}.

Multi-phase field models represent each cell as a field. In comparison to vertex models, they therefore allow for a more detailed description of cell shape at the cost of being more computationally expensive. We introduce junctional fluctuations by modifying the cell-cell adhesion term in the model, with pairwise adhesions now following an Ornstein-Uhlenbeck process as has been done in vertex models. We show that, for appropriate values of the adhesion fluctuation variance and persistence, these fluctuations can indeed induce rearrangements and fluidise the multi-phase field model. Interestingly, we find the same non-monotonic dependence of the cells' effective diffusion coefficient that has previously been observed in the vertex model~\cite{yamamoto:2022}.

%||||||||||||||||||||||||||||||||||||||||||||||||||||||||||||
%||||||||||||||||||||||||||||||||||||||||||||||||||||||||||||
%||||||||||||||||||||||MODEL|||||||||||||||||||||||||||||||||
%||||||||||||||||||||||||||||||||||||||||||||||||||||||||||||
%||||||||||||||||||||||||||||||||||||||||||||||||||||||||||||
\section{Phase-Field Model}

In the multi-phase field model, each cell $i$ is represented by a phase field $\varphi^{(i)}(\mathbf{x},t)$~\cite{mueller:2019}, which takes value $1$ inside and $0$ outside the cell. The dynamics of the phase fields are given by the equation of motion
\begin{equation}
    \label{eq:dynamics}
    \partial_t \varphi^{(i)} + \mathbf{v}^{(i)} \cdot \nabla \varphi^{(i)}= -J_0 \frac{\delta {\mathcal{F}}}{\delta \varphi^{(i)}},
\end{equation}
where $\mathbf{v}^{(i)}(\mathbf{x},t)$ is the velocity of the phase field $\varphi^{(i)}(\mathbf{x},t)$, $\mathcal{F}$ is the free energy of the tissue, and the right hand side represents the relaxation to a free energy minimum at a rate $J_0$. The velocity $\mathbf{v}^{(i)} (\mathbf{x},t)$ is determined in the overdamped regime by the local force density acting on the cell,
\begin{equation}
    \label{eq:force_balance}
    \xi \mathbf{v}^{(i)} (\mathbf{x},t) =\mathbf{f}^{(i)} (\mathbf{x},t),
\end{equation}
where $\xi$ is a friction coefficient and the force density is
\begin{equation}\label{eqn:passive}
    \mathbf{f}^{(i)}(\mathbf{x},t)=\frac{\delta\mathcal{F}}{\delta\varphi^{(i)}}\nabla\varphi^{(i)}.
\end{equation}
The free energy $\mathcal{F}=\mathcal{F}_{CH}+\mathcal{F}_{area}+\mathcal{F}_{rep}+\mathcal{F}_{adh}$ has four contributions~\cite{mueller:2019,zhang:2020,mueller:2021}:
\begin{subequations}\label{eqn:f}
    \begin{align}
        \mathcal{F}_{CH}&=\sum_i\frac{\gamma}{\lambda}\int\textrm{d}\mathbf{x}\Big[\varphi^{(i)}(1-\varphi^{(i)})^2+\lambda^2(\nabla\varphi^{(i)})^2\Big]\label{ch},
        \\
        \mathcal{F}_{area}&=\sum_i\mu\left[1-\frac{1}{\pi R^2}\int\textrm{d}\mathbf{x}\,(\varphi^{(i)})^2\right]^2\label{area},\\
        \mathcal{F}_{rep}&=\sum_i\sum_{j\neq i}\frac{\kappa}{\lambda}\int\textrm{d}\mathbf{x}\,(\varphi^{(i)})^2(\varphi^{(j)})^2\label{rep},\\
        \mathcal{F}_{adh}&=\sum_i\sum_{j\in \mathcal{N}_i(t)}\omega_{ij}\lambda\int\textrm{d}\mathbf{x}\,\nabla\big[(\varphi^{(i)})^2\big]\cdot\nabla\big[(\varphi^{(j)})^2\big]\label{adh}.
    \end{align}
\end{subequations}
Of these, the first is a Cahn-Hilliard term favouring phase separation into regions of $\varphi^{(i)}=0,1$. This results in an interface of width $\mathcal{O}(\lambda)$ lattice units, interpreted as the cell membrane, with $\gamma/\lambda$ being an energy scale. The second term is a soft area constraint with strength $\mu$ and equilibrium cell area $\pi R^2$ given by the radius $R$. The third term imposes a penalty on cell-cell overlap with an energy scale $\kappa/\lambda$.

The fourth term in the free energy describes the adhesion of cell membranes by computing the dot product of the gradients of the squares of each pair of phase fields. The inside (outside) of a cell corresponds to $\varphi=1$ ($\varphi=0$) so when two phase-field cells are next to each other, the gradients of their squares point in approximately opposite directions. The strength of the adhesion between cells $i$ and $j$ is given by $\omega_{ij}$, and it is energetically favourable (unfavourable) for the interfaces of cells $i$ and $j$ to overlap when $\omega_{ij}$ is positive (negative). The total adhesive free energy is computed as a pairwise sum over all neighbours, where $\mathcal{N}_i(t)$ denotes the set of neighbours of cell $i$ at time $t$. This term broadly corresponds to the line tension term in vertex models~\cite{farhadifar2007influence,krajnc:2020}, which captures both tension and cell-cell adhesion.

This adhesion free energy functional differs from those of some previous works~\cite{zhang:2020,zhang:2021} in two key ways. Firstly, in this formulation, the functional derivative of $\mathcal{F}_{adh}$ that appears on the right-hand side of Eq.~\eqref{eq:dynamics} is proportional to $\varphi^{(i)}$. A previous implementation of adhesion in the multi-phase field model \cite{zhang:2020} does not have this proportionality and can exhibit numerical instability when $\omega$ is large (see Supplemental Material~\cite{SI}). Secondly, other works have considered a single energy scale $\omega$ for all adhesive interactions, while here we implement a set of pairwise $\{\omega_{ij}\}$ so that each pair of cells can have its own adhesion strength.

These $\omega_{ij}$ are the avenue through which we introduce junctional fluctuations to the model, representing the stochastic turnover of junctional molecular motors. We demand symmetry in the energy scales, $\omega_{ij}=\omega_{ji}$, so that the pairwise interactions of neighbour cells are reciprocal. The values of $\omega_{ij}$ follow an Ornstein-Uhlenbeck process, as had previously been done in vertex models $\cite{curran:2017,krajnc:2020,duclut:2021a,yamamoto:2022}$:
\begin{equation}
    \frac{\textrm{d}\omega_{ij}(t)}{\textrm{d}t}=-\frac{1}{\tau_{\omega}}\omega_{ij}(t)+\xi_{ij}(t),
    \label{eqn:fluctuations}
\end{equation}
where $\xi_{ij}$ is Gaussian white noise with properties $\langle\xi_{ij}(t)\rangle=0$ and $\langle\xi_{ij}(t)\xi_{kl}(t')\rangle=\left(2\sigma^2_\omega/\tau_\omega\right)\delta_{ik}\delta_{jl}\delta(t-t')$, whereas $\tau_{\omega}$ is the persistence time and $\sigma^2_\omega$ is the long-time variance of the adhesion fluctuations. Because the strength of adhesion is only meaningful between adjacent cells, $\omega_{ij}(t)$ is only evolved with time if cells $i,j$ are neighbours; otherwise, it is set to $0$. This means that when a pair of cells newly come into contacts their adhesion is zero.

We simulate $100$ cells in a rectangular domain with side lengths $L_x=150$ and $L_y=130$ lattice units and periodic boundary conditions. The cells are initialised in a regular hexagonal geometry. We allow the system to run for $3\times10^4$ time units before taking measurements. Time $t=0$ corresponds to the start of measurements, and the simulation then runs until time $t=T=1.8\times10^5$. Data about the state of the tissue is output in intervals of length $\delta t=1000$.

The parameters assigned to the free energy terms in Eq.~\eqref{eqn:f} are $\gamma=1.4$, $\lambda=2.0$, $\mu=120$, and $\kappa=1.5$. The coefficient of friction in Eq.~\eqref{eq:force_balance} is $\xi=3.0$ and the target cell area is given by radius $R=8.0$ (in lattice units). Relaxation according to Eq.~\eqref{eq:dynamics} is controlled by the rate $J_0=10^{-2}$. The phase-field Eqs.~\eqref{eq:dynamics}-\eqref{eqn:f} are solved using a one-step predictor-corrector method. The unit of time is 1, and each unit is divided into five timesteps. The method of solving the phase-field model is outlined further in \cite{mueller:2019,mueller:2021}. The adhesion fluctuations of Eq.~\eqref{eqn:fluctuations} are solved using a finite difference method on each timestep.

%||||||||||||||||||||||||||||||||||||||||||||||||||||||||||||
%||||||||||||||||||||||||||||||||||||||||||||||||||||||||||||
%||||||||||||||||||||||RESULTS|||||||||||||||||||||||||||||||
%||||||||||||||||||||||||||||||||||||||||||||||||||||||||||||
%||||||||||||||||||||||||||||||||||||||||||||||||||||||||||||
\section{Results}

An epithelial monolayer can change its topology through several processes, such as cell division and extrusions \cite{saw:2017,merkel:2017,lin:2022,monfared:2023}. Fluidisation of a tissue through fluctuating tensions/adhesions at cell-cell junctions relies on cell intercalations (i.e., T1 transitions) that are able to modify neighbour-neighbour contacts \cite{merkel:2017,duclut:2021,sknepnek:2022}. During a T1 transition, the junction between one pair of cells shrinks and a new junction forms between another previously unconnected pair of cells. Each of the two cells that were neighbours before the transition loses one neighbour, while each of the two cells that are neighbours after the transition gains one.

\begin{figure}[h]
    \centering
    \includegraphics[width=8.6cm]{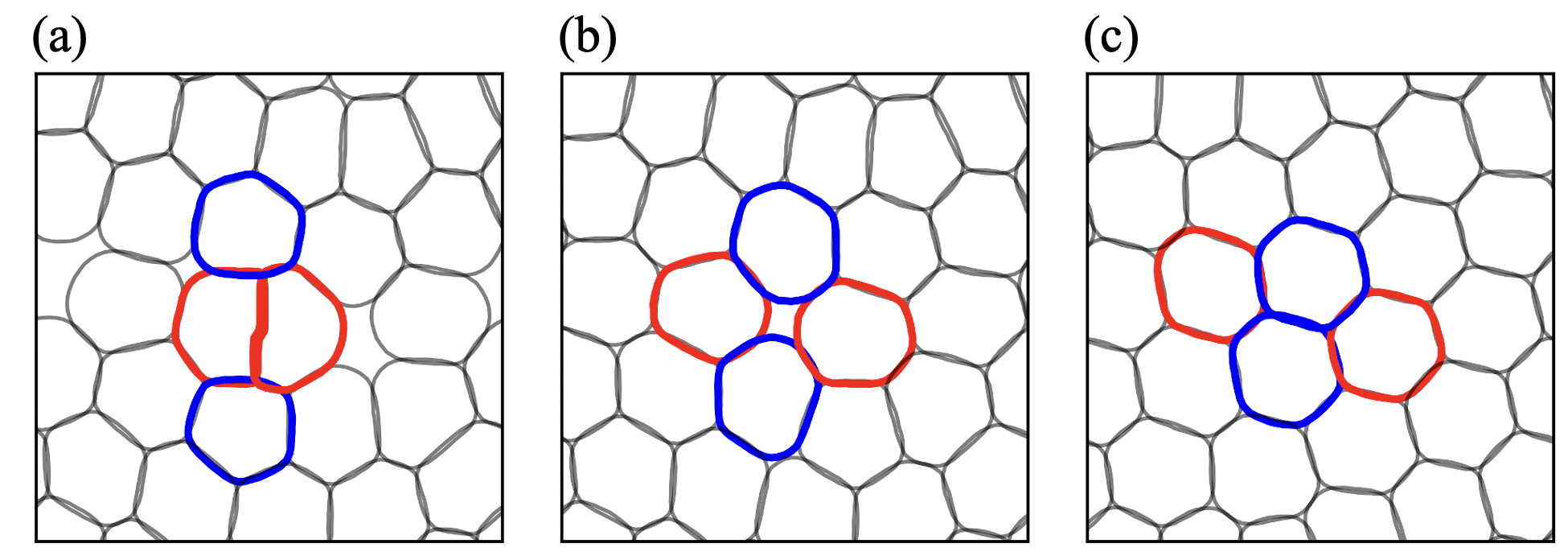}
    \caption[A quartet of cells before, during, and after a T1 topological transition]{A quartet of cells (a) before, (b) during, and (c) after a T1 topological transition in an adhesive background. The red cells at first attract with $\omega_{ij}>0$ (a) and then repel with $\omega_{ij}<0$ (b), while the blue cells attract with $\omega_{ij}>0$ to complete the transition (c).}
    \label{fig:T1}
\end{figure}

We first show that a T1 can be induced in the multi-phase field model by modulating the pairwise cell adhesions $\omega_{ij}$ (Fig.~\ref{fig:T1} and Movie S1). In Fig.~\ref{fig:T1}(a), the cells outlined in red are stuck together by a large positive $\omega_{ij}=+10$, whereas all other pairwise adhesions are given by $\omega_{ij}=0.4$. The adhesion parameter for the red cells is then set to $-10$ so that the cells are pushed apart in Fig.~\ref{fig:T1}(b). The cells transition through the equivalent of a fourfold vertex until the cell intercalation is completed in Fig.~\ref{fig:T1}(c). So that the system does not arrest before completing the T1, we add a noise term $D_\varphi\eta$ to the RHS of Eq.~\eqref{eq:dynamics} for this simulation, where $\eta$ is unit-variance Gaussian noise and $D_\varphi=0.1$.

%|||||||||||||||||||||||||||||||||||||||||
%|||||||||||||FIGURE 1|||||||||||||||||||| 
%|||||||||||||||||||||||||||||||||||||||||

\begin{figure}[h!]
    \centering
    \includegraphics{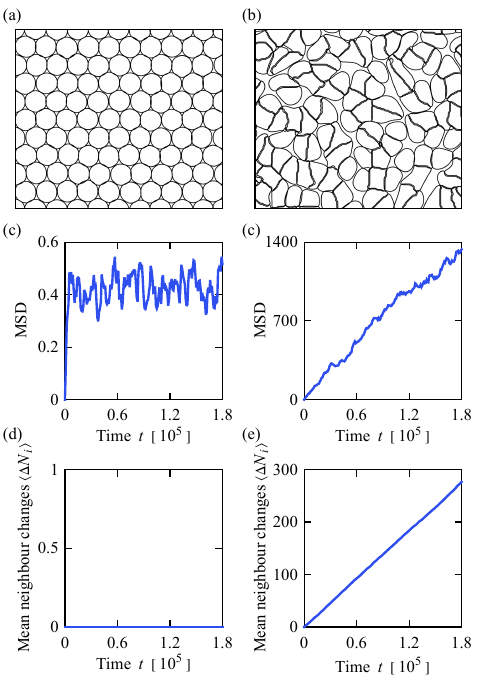}
    \caption{Model tissue with (a) $\sigma_\omega=0.1$, $\tau_{\omega}=10^{3}$ and (b) $\sigma_\omega=1.1$, $\tau_\omega=10^3$. The larger $\sigma_\omega$ in (b) permits greater variance in $\omega_{ij}$, which result in intercalations and diffusive behaviour. Note that the long contacts of some cell pairs arise from high adhesion between those cells. (c,d) Mean-square displacement plotted against time for the systems in (a) and (b), respectively. (e,f) Average neighbour changes plotted against time for the systems in (a) and (b), respectively. The solid phase exhibits no neighbour changes whereas the fluid phase has a constant rate of neighbour turnover.}
    \label{fig:msd}
\end{figure}

We next turn to simulating a tissue with all pairwise adhesions fluctuating following Eq.~\eqref{eqn:fluctuations} (without the additional $D_\varphi\eta$ term in Eq.~\eqref{eq:dynamics}). We first set $\sigma_\omega=0.1$ and $\tau_\omega=10^3$. The resulting monolayer is shown in Fig.~\ref{fig:msd}(a) and Movie S2: the system, despite fluctuations in $\omega_{ij}$, is solid-like (i.e. cells do not exchange neighbours) and remains in a regular hexagonal lattice. 

Keeping $\tau_\omega=10^3$, we next perform a simulation at a higher $\sigma_\omega=1.1$, with the resulting model tissue shown in Fig.~\ref{fig:msd}(b) and Movie S3. The larger fluctuations in adhesion resulting from a higher $\sigma_\omega$ can now overcome the energy barriers associated with neighbour exchanges, enabling cells to escape from their local neighbourhood through a series of T1 transitions and become mobile, producing a disordered, fluid-like tissue.

To quantify the difference in mobility between the two cases, we measure the mean-square displacement
\begin{equation}\label{eqn:msd-ch6}
    \text{MSD}(t)=\frac{1}{N}\sum_{i=1}^{N}\left|\mathbf{r}^{(i)}(t)-\mathbf{r}^{(i)}(0)\right|^2,
\end{equation}
where $N$ is the number of cells and $\mathbf{r}^{(i)}$ is the centre-of-mass position of cell $i$ at time $t$. As expected, in the solid-like model tissue, MSD saturates at a small value (less than the cell radius $R$), as shown in Fig.~\ref{fig:msd}(a). In a fluidised layer, however, the MSD grows linearly in time [Fig.~\ref{fig:msd}(b)]. 

As a second measure of fluidisation, we also estimate how often cells gain and lose neighbours. Specifically, we compare how the set of neighbours $\mathcal{N}_i(t)$ of cell $i$ changes between subsequent outputs of the tissue state, which are separated in time by $\delta t=1000$. For each neighbouring cell that that is either added or removed from $\mathcal{N}_i(t)$, the number of the cell's neighbour change events $\Delta N_i$ is incremented by $1$; as the tissue states are only compared every $\delta t$, this does not account for every change of the neighbour structure, but allows for a comparison of such events between different simulations. Note that a cell can separate and reconnect with another cell multiple times while remaining in the same general neighbourhood.

%|||||||||||||FIGURE 2||||||||||||||||||||
%|||||||||||||||||||||||||||||||||||||||||

%|||||||||||||||||||||||||||||||||||||||||
%|||||||||||||FIGURE 3||||||||||||||||||||
%|||||||||||||||||||||||||||||||||||||||||
\begin{figure*}
    \centering
    \includegraphics{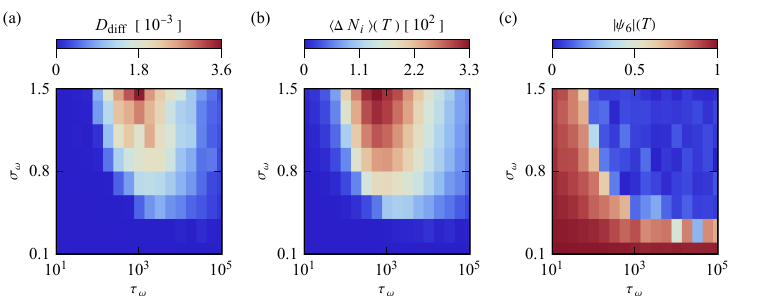}
    \caption{Heatmaps of (a) the effective diffusion coefficient $D_\textrm{diff}$, (b) the final average estimated number of neighbour change events $\left<\Delta N_i\right>$, and (c) final $|{\psi_6}|$ in $\tau_\omega$ -- $\sigma_\omega$ phase space. Panels a and b show that the effective diffusion coefficient and the number of neighbour change events depend non-monotonically on $\tau_\omega$. Panel c illustrates the change from high to low $|{\psi_6}|$, indicating, respectively, a regular hexagonal lattice
     and a disordered state.}
    \label{fig:heatmaps}
\end{figure*}

The resulting plots of $\Delta N_i$ averaged over all cells are shown in Fig.~\ref{fig:msd}(e),(f). There are no neighbour changes in the solid-like tissue [Fig.~\ref{fig:msd}(e)] but the average number of changes grows linearly in the fluid-like tissue [Fig.~\ref{fig:msd}(f)], revealing a constant rate of neighbour exchange. The presence of neighbour changes, in conjunction with the diffusive MSD, discounts solid flocking~\cite{zhang:2020} as the origin of the measured cell mobility. 

We then perform a scan of $\tau_\omega$ -- $\sigma_\omega$ parameter space and characterise the resulting monolayers using the effective diffusion coefficient, the frequency of neighbour-change events, and the hexatic order parameter. These quantities are illustrated in heatmaps in Fig.~\ref{fig:heatmaps}. The effective diffusion coefficient is calculated as $D_\textrm{diff}\equiv \text{MSD}(T)/(4T)$, where $T=1.8\times 10^5$ is the final simulation time. At a given $\tau_\omega$, the effective diffusion coefficient $D_\textrm{diff}$ increases with $\sigma_\omega$. However, we find that $D_\textrm{diff}$ depends non-monotonically on $\tau_\omega$ [Fig.~\ref{fig:heatmaps}(a)]. This is in agreement with the observations of Yamamoto et al. in the vertex model~\cite{yamamoto:2022}.

We then plot how the final value at $t=T$ of estimated neighbour change events averaged over all cells $\left<\Delta N_i\right>$ depends on $\sigma_\omega$ and $\tau_\omega$ in Fig.~\ref{fig:heatmaps}(b). A solid phase should exhibit no neighbour rearrangements following any initial annealing of imperfections, whereas cells in a fluid phase should change neighbours regularly. Indeed, we observe that for sufficiently large $\tau_\omega$ and $\sigma_\omega$, neighbour changes occur at finite rate. Just as for the effective diffusion coefficient, the dependence on $\tau_\omega$ is non-monotonic, so that there is an optimal $\tau_\omega$ that maximises the number of neighbour change events.

Finally, the hexatic order of the model tissue is quantified using the hexatic order parameter $\psi_6$, famously a feature of the KTHNY theory of crystal melting in two dimensions \cite{halperin:1978,nelson:1979}, that has previously been used to characterise structure in model tissues~\cite{li2018role,pasupalak2020hexatic,lin:2022,keta2025long}. It is calculated as
\begin{equation}\label{eqn:psi6}
    {\psi_6}_j=\frac{1}{N_j}\sum_{k\in \mathcal{N}_j}\exp(6i\theta_{k,j})
\end{equation}
where $j,k$ index cells, $\mathcal{N}_j$ is the set of neighbours of cell $j$, $N_j$ is the size of said set, and $\theta_{k,j}$ gives the angle of the centre of mass of cell $k$ with respect to that of cell $j$. We measure $|{\psi_6}|\equiv|{\langle{\psi_6}_j\rangle}|$ the magnitude of the hexatic structure across all cells in a layer. As the initial condition is a regular hexagonal lattice, $|{\psi_6}|$ should take a value close to $1$ for a solid-like state and close to $0$ for a fluid-like state. A heatmap of final $|{\psi_6}|$ is shown in Fig.~\ref{fig:heatmaps}(c). As expected, the mean hexatic order parameter changes from 1 to 0 as $\tau_\omega$ and $\sigma_\omega$ become sufficiently large, indicative of a transition from a regular hexagonal lattice to a disordered state.

Lastly, note that as the simulations are relatively computationally expensive, the final simulation time is comparable with the highest values of $\tau_\omega$ we consider. However, the non-monotonic behaviour of $D_\textrm{diff}$ and final $\left<\Delta N_i\right>$ can already be observed for $\tau_\omega<10^4$, so at least an order of magnitude smaller than the simulation time (Fig.~\sidiff~\cite{SI}). Moreover, at $\tau_\omega=10^4$ and even $\tau_\omega=10^5$, the MSD plots are still approximately diffusive on the simulation time scale (Fig.~\sifit~\cite{SI}); the final MSD at $\tau_\omega=10^4$ and $\sigma_\omega=1.5$ also corresponds to cell displacement by more than two cell diameters. Together, these suggest that the non-monotonic dependence on $\tau_\omega$ is a property of the system and not a consequence of the simulation length.

%||||||||||||||||||||||||||||||||||||||||||||||||||||||||||||
%||||||||||||||||||||||||||||||||||||||||||||||||||||||||||||
%||||||||||||||||||||||SUMMARY|||||||||||||||||||||||||||||||
%||||||||||||||||||||||||||||||||||||||||||||||||||||||||||||
%||||||||||||||||||||||||||||||||||||||||||||||||||||||||||||
\section{Discussion}
We studied a multi-phase field model for an epithelium~\cite{mueller:2019,zhang:2020} augmented with fluctuating pairwise adhesions following an Ornstein-Uhlenbeck process, representing the stochastic turnover of junctional motors. We showed that adhesion fluctuations result in rearrangements leading to diffusive motion of cells in the model monolayer. This provides another fluidisation mechanism for phase-field models, focusing on the role of cell-cell junctions, rather than cell-scale  polar~\cite{zhang:2020,loewe:2020} or nematic~\cite{mueller:2019,zhang:2021} activity.

Such Ornstein-Uhlenbeck fluctuations have previously been studied as a fluidisation mechanism in vertex models \cite{curran:2017,krajnc:2020,duclut:2021,yamamoto:2022}. Our work validates these results for a different type of discrete cell-level model. Moreover, we find the cells' effective  diffusion coefficient to have a non-monotonic dependence on persistence time in adhesion fluctuation, in agreement with what was reported in the vertex model~\cite{yamamoto:2022}. That study suggested the non-monotonic dependence arises from a separation of timescales at large fluctuation persistence times, with the system reaching a nearly force balanced state quickly after a rearrangement and that state then becoming unstable again on the timescale of fluctuation persistence.

Given the considerable range of uses the fluctuating-tensions approach found in the vertex model literature, we hope that the model proposed here will also be a powerful tool for further studies using multi-phase field models. In future work, pairwise adhesion terms could be modified to, e.g., also account for the length of cell-cell interfaces~\cite{zankoc:2020,krajnc:2021}, respond to local tensions~\cite{sknepnek:2022}, or depend on cell shape~\cite{rozman:2023}.

\begin{acknowledgments}
We wish to thank Ioannis Hadjifrangiskou, Matej Krajnc, Gianmarco Spera, and Julia M. Yeomans for helpful discussion. We also wish to thank Matej Krajnc and Julia M. Yeomans for a careful reading of the manuscript. J.N.G. acknowledges Pembroke College, Oxford. J.R. acknowledges support from the UK EPSRC (Award EP/W023849/1) and ERC Advanced Grant ActBio (funded as UKRI Frontier Research Grant EP/Y033981/1).
\end{acknowledgments}

\end{document}

\newcommand{\zc}{\zeta_{\rm{c}}}
\newcommand{\ze}{\zeta_{\rm{e}}}

\newcommand{\figdroplet}{3}

\preprint{}

\title{
\textbf{Supplemental Material: \\ Junctional-Fluctuation-Mediated Fluidisation of \\ Multi-Phase Field Epithelial Monolayers}}
\author{James N. Graham}
\affiliation{Rudolf Peierls Centre for Theoretical Physics, University of Oxford, Oxford OX1 3PU, United Kingdom}
\author{Jan Rozman}
\affiliation{Rudolf Peierls Centre for Theoretical Physics, University of Oxford, Oxford OX1 3PU, United Kingdom}

\date{\today}

%\keywords{Suggested keywords}%Use showkeys class option if keyword
                              %display desired
{
\let\clearpage\relax
\maketitle
}

\section*{Adhesion free energy functional}

The free energy functional for adhesion in this work is given by
\begin{equation}
\mathcal{F}_{adh}=\sum_i\sum_{j\in \mathcal{N}_i(t)}\omega_{ij}\lambda\int\textrm{d}\mathbf{x}\,\nabla\big[(\varphi^{(i)})^2\big]\cdot\nabla\big[(\varphi^{(j)})^2\big].\label{fnc:new}
\end{equation}
Assuming $\varphi^{(i)}\to0$ at long distances gives
\begin{equation}
\frac{\delta {\mathcal{F}_{adh}}}{\delta \varphi^{(i)}}=-4\sum_{j\in \mathcal{N}_i(t)}\omega_{ij}\lambda\,\varphi^{(i)}\nabla^2[(\varphi^{(j)})^2].
\end{equation}
This is in contrast to a previous formulation of adhesion in the phase field model given by \cite{zhang:2020}
\begin{equation}
\hat{\mathcal{F}}_{adh}=\sum_i\sum_{j\in \mathcal{N}_i(t)}\omega\lambda\int\textrm{d}\mathbf{x}\nabla\varphi^{(i)}\cdot\nabla\varphi^{(j)},\label{fnc:old}
\end{equation}
which has a functional derivative
\begin{equation}
\frac{\delta {\hat{\mathcal{F}}_{adh}}}{\delta \varphi^{(i)}}=-2\sum_{j\in \mathcal{N}_i(t)}\omega\lambda\,\nabla^2\varphi^{(j)}\label{adh:old}.
\end{equation}
Therefore, the modification to the adhesion term results in the functional derivative being proportional to $\varphi^{(i)}$, making it less likely that it has a large value in regions of low $\varphi^{(i)}$ (i.e., far from the bulk of cell $i$). We find that this improves the stability of the model when $\omega_{ij}$ is large.

\section*{Supplementary figures}
\begin{figure}[h]
    \centering
    \includegraphics[width=8 cm]{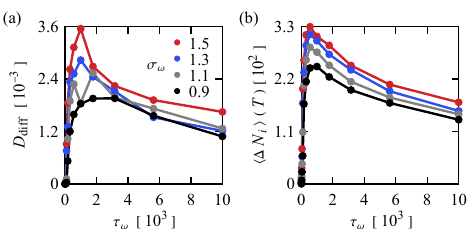}
    \caption{Effective diffusion coefficient (a) and estimated final number of neighbour change events averaged over all cells (b) as a function of $\tau_\omega$ at different $\sigma_\omega$, zooming in on $\tau_\omega\leq10^4$. Legend on panel (a) applies to both panels.}
    \label{sfig:diff}
\end{figure}

\begin{figure}[h]
    \centering
    \includegraphics[width=8 cm]{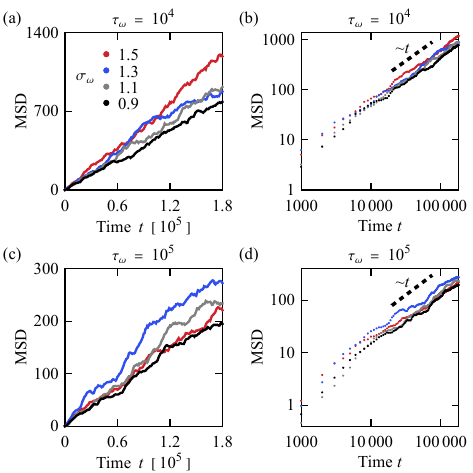}
    \caption{(a,b) MSD as a function of time for different $\sigma_\omega$ at $\tau_\omega=10^4$ shown on a linear (a) and a log-log (b) scale. Exponents of a $\textrm{MSD}(t)=\alpha t^\beta$ fit for $\sigma_\omega=1.5$, $1.3$, $1.1$, and $0.9$ are, respectively, $1.09$, $0.86$, $1.04$, and $1.03$. (c,d) MSD as a function of time for different $\sigma_\omega$ at $\tau_\omega=10^5$ shown on a linear (c) and a log-log (d) scale. Exponents of a $\textrm{MSD}(t)=\alpha t^\beta$ fit for $\sigma_\omega=1.5$, $1.3$, $1.1$, and $0.9$ are, respectively, $0.94$, $0.87$, $1.03$, and $0.97$. Legend on panel (a) applies to all panels.}
    \label{sfig:fit}
\end{figure}

\section*{Movie captions}
\begin{itemize}
    \item \textbf{Movie S1:} T1 transition induced in the multi-phase field model: red cells initially have pairwise adhesion $\omega_{ij}=10$, whereas all other pairwise adhesions are $\omega_{ij}=0.4$. The adhesion between the red cells is then changed to $-10$, producing a T1 transition.
    \item \textbf{Movie S2:} Multi-phase field model tissue evolution for $\sigma_\omega=0.1$ and $\tau_\omega=10^3$.
    \item \textbf{Movie S3:} Multi-phase field model tissue evolution for $\sigma_\omega=1.1$ and $\tau_\omega=10^3$. One cell (red) and its neighbours at $t=0$ are coloured to enable tracking.
\end{itemize}